\documentclass[letterpaper]{jpconf}
\usepackage{graphicx}
\begin{document}
\title{Update on {\boldmath $Z(4430)$} and {\boldmath $X(3872)$} at Belle}

\author{Sadaharu Uehara  (for the Belle Collaboration)}

\address{KEK-IPNS, Institute of Nuclear and Particle Studies, High Energy Accelerator Research Organization, Tsukuba 305-0801 Japan}

\begin{abstract}
The $X(3872)$ and $Z(4430)$ are candidates of tetraquark state with a $c\bar{c}$ pair. We present results from Belle recently updated  for
the mass, branching fractions etc. in different production/decay processes of the
$X(3872)$. Results from a Dalitz analysis for $B \to Z(4430) K \to \psi' \pi^\pm K$ are also presented.

\end{abstract}
 
\section{Introduction}
We face to a flood of discoveries of new charmonium-like states brought from the B-factory experiments. Some of these states seem to be attributed to a sequence
of the ordinary charmonium states, but true nature of many of them are 
still unknown. 
If some of them have no correspondence to any empty seats in the sequential states or  
have any peculiar properties different from ordinary charmonium, the states 
must be considered as candidate of exotic states.  

 The B-factory experiments have found these charmonium-like states in different processes:
$B$-decay processes, single-photon annihilations mainly as the ISR processes, double charmonium production processes and two-photon processes. 
Among these states, we here present on  $X(3872)$ and $Z(4430)$, 
which are serious candidates of the exotic states found in $B$ decays.  
The measurements reported here have been performed using Belle detector~\cite{belle} at the KEKB accelerator~\cite{kekb}. 

\section{Known features of {\boldmath $X(3872)$}}  
$X(3872)$ was discovered by the Belle experiment in 2003 in the decay process of $B$ meson, $B^0 \to J/\psi \pi^+\pi^-$~\cite{xdisc}. The width was very narrow, and that was argued with a relation to its possible exotic natures.  Slightly later, this state was confirmed by BaBar and two Tevatron experiments. Measurements of the width until the present shows a narrower value than the 
measurement limit of any experiments.

  After the discovery, some important and interesting natures of $X(3872)$ are clarified. The charge conjugation is determined to be positive from a presence of the $\gamma J/\psi$ decay mode, and in contrast the $\gamma \chi_{cJ}$ decays are not observed. The invariant mass distribution of the $\pi^+ \pi^-$ system shows a 
$\rho$-meson like nature. The invariant-mass distribution for another final state, to $J/\psi$ and three pions, shows a rise toward the $\omega$ mass.  This is an indication of isospin non-conservation in decays of $X(3872)$. In addition, from studies of decay angles, a probable spin-parity ($J^P$) assignment of $1^+$ is suggested~\cite{cdfx}.

\section{Search for mass-splitting of {\boldmath $X(3872)$}}
  We show the results updated by Belle in the summer 2008, which are preliminary. Some theoretical models treating $X(3872)$ as a two-meson molecule state predict different branching fractions or a mass splitting in the decays to $X(3872)$ from the different charges of $B$ mesons, that is,  some differences between the $B^0 \to X(3872) K^0$ and $B^\pm \to X(3872) K^+$ decays.  

 New measurement from Belle of the ratio of the branching fractions from 
$B^0$ and $B^\pm$ is 
${\cal B}(B^0 \to X K^0)/{\cal B}(B^\pm \to X K^\pm)=0.82 \pm 0.22 \pm 0.05$, and this ratio is consistent with unity~\cite{bellex1}.  Moreover, 
we do not find any significant difference between the masses of the $X(3872)$ 
in the two decay modes, $M(X {\rm \ from\ } B^\pm)-M(X {\rm \ from\ } B^0) = +0.18 
\pm 0.89 \pm 0.26$~MeV/$c^2$.  We do not observe any mass splitting.

  Through this analysis, Belle obtains the updated measurement of the mass, $M(X(3872))= 3871.46 \pm 0.37 \pm 0.07$~MeV/$c^2$. It should be remarkable that the mass still coincides with the sum of the $D^0$ and $D^{*0}$ masses within the error of about 0.4~MeV/$c^2$.  

\begin{figure}
\centering
\includegraphics[width=11cm]{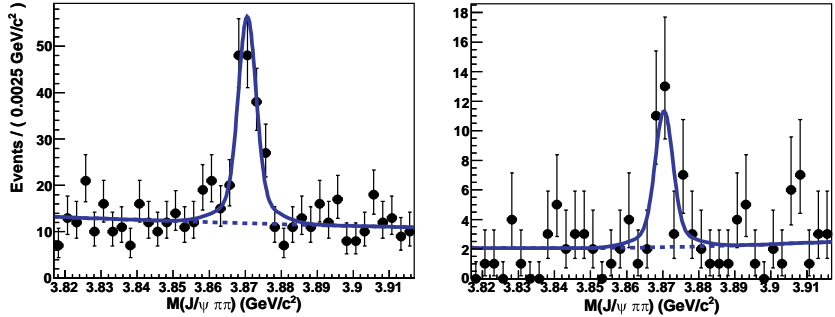}
\label{fig1}
\centering
\caption{Invariant-mass distribution of $J/\psi \pi^+ \pi^-$ from
the $B$ decays associated by a charged kaon (left) and a neutral kaon
(right).}
\end{figure}

\section{Other production and decay modes of {\boldmath $X(3872)$}}
Belle has found a new production mode of $X(3872)$ from $B$ decays being associated by a $K \pi$ pair (Fig.~2(left))~\cite{bellex1}.  The study of the $K \pi$ invariant mass spectrum shows that $K^*(892)$ component is not significant (Fig.~2(right)). The small peak in the spectrum dominantly comes from the non-$X(3872)$ component, and the contribution of the $K^*$ peak associating  $X(3872)$ seems to be small even if it exists. This is in contrast to $J/\psi$ or $\psi(2S)$ production in the similar decays of $B$ mesons where the associations of 
$K \pi$ are dominated by $K^*(892)$.

  The decay mode to the $D^0 \bar{D^0} \pi^0$ final state is another important branch of 
$X(3872)$. Belle found a threshold enhancement in this final state 
from the $B$ decays associated by a kaon~\cite{belledds}. The mass 
corresponding to about 3875~MeV/$c^2$, a little heavier than the nominal 
$X(3872)$ mass, was reported.
BaBar also observed a similar enhancement with its heavier mass~\cite{babardds}.

  Belle has updated the analysis of this channel, with assuming an intermediate $D^{*0}$ state 
decaying to the $D^0 \pi^0$ and $D^0 \gamma$ with the known branching-fraction ratio. 
The updated mass result, $3872.6^{+0.5}_{-0.4} \pm 0.4$~MeV/$c^2$, is now close to the nominal mass of $X(3872)$.  Figure~3 shows the reconstructed $D^0 \bar{D}^{*0}$ mass spectra for the two decay modes.

\begin{figure}
\centering
\includegraphics[width=11cm]{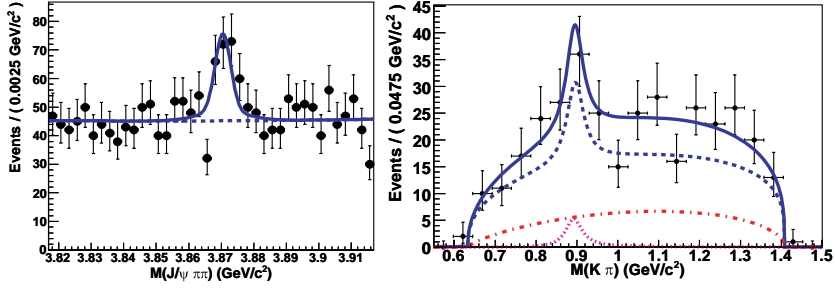}
\label{fig2}
\centering
\caption{Invariant-mass distribution of $J/\psi \pi^+ \pi^-$ from
the $B$ decays associated by a $K \pi$ pair (left) and
the invariant-mass distribution of the $K \pi$ with the fit curves(right):
non-resonant $K \pi$ component is shown by the dash-dot red curve, 
$X(3872)K^*(892)^0$ by the dotted
magenta curve, and the background by the dashed blue curve.}
\end{figure}

\begin{figure}
\centering
\includegraphics[width=11cm]{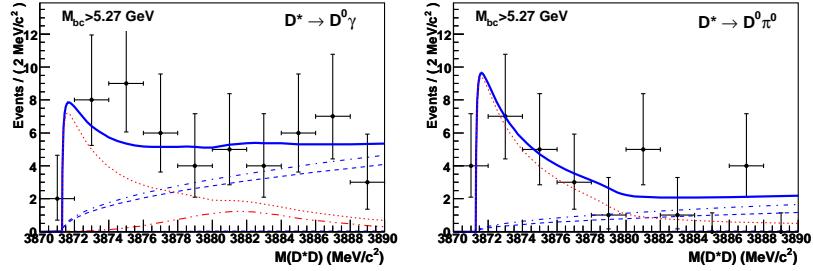}
\label{fig3}
\centering
\caption{ Invariant-mass distributions of $D^0 \bar{D}^{*0}$ with
the two decay modes of $\bar{D}^{*0}$ to $\bar{D}^0 \gamma$ (left) and 
$\bar{D}^0 \pi^0$ (right). The dotted and dashed curves are the resonance and
the background distributions from the fit, respectively.}
\end{figure}

\section{Dalitz analysis of {\boldmath $B \to Z(4430) K$, $Z(4430) \to \psi' \pi^\pm$} }
  $Z(4430)$ is discovered in the decay of $B$ mesons to $\psi'$, 
a charged pion and a kaon, by Belle~\cite{zdisc}. We find an enhanced horizontal 
band at around $M(\psi' \pi)^2 =17$~GeV$^2$ in the Dalitz plot of the three-body decays to $\psi' \pi^\pm K$ (Fig.~4(left)).  
At the same time, we find two vertical bands from the $K \pi$ resonances, $K^*(892)$ and 
$K^*(1430)$. Belle reported the observation of the new state with a $6.5 \sigma$ significance in the $\psi' \pi$ mass distribution after vetoing the events in these two $K^*$ regions. Because this is a charmonium-like state with non-zero electric charge, 
it can no longer be a simple $c\bar{c}$ state, and this must 
be a very serious tetraquark candidate.

  Later, the BaBar collaboration tried to search for the same state in their data~\cite{babarz}. They analyzed the events in the full Dalitz plane introducing some $K^*$ sates, and  they claimed that they did not find a significant peak from $Z$. The difference between the 
BaBar and Belle results is of $1.7 \sigma$. 
Thus, there is no conclusive evidence of $Z(4430)$ from the BaBar result.

  Belle also has tried a Dalitz analysis introducing many $K \pi$ resonances, using the same event sample as in the previous publication~\cite{bellez2}. First, we tried the fit without including the $Z(4430)$ but with introducing 6 kaon-resonance states,
$\kappa$ or $K^*(800)$, $K^*(892)$, $K^*(1410)$, $K^*_0(1430)$,  $K^*_2(1430)$ and $K^*(1680)$ ($K^*_3(1780)$ was finally removed from the standard fit). 
The fit could not reproduce the distribution, well.  Especially, any interferences among the $K^*$ resonances seems to unable to reproduce the peak at the $Z(4430)$ mass location. With an inclusion of $Z(4430)$, the fit is drastically improved, and the presence of the $Z(4430)$ signal is confirmed by a $6.4 \sigma$ significance.  No characteristic feature of the $K \pi$ resonance shapes is seen in  the $Z$ resonance region of the Dalitz slice (Fig.~4(middle)) . 

 Figure~4(right) shows the fits with and without assuming the $Z$ resonance summed in
the $K \pi$ mass regions where $K^*(892)$  and $K^*_{0,2}(1430)$ are vetoed.
The resonance parameters of $Z(4430)$  obtained by this analysis are $M = 4443^{+15~+19}_{-12~-13}$~MeV/$c^2$ and $\Gamma = 107^{+86~+74}_{-43~-56}$~MeV, 
where $M$ and $\Gamma$ are the mass and total width of $Z(4430)$, respectively.
The product of the branching fractions, ${\cal B}(\bar{B}^0 \to K^- Z^+){\cal B}(Z^+ \to \pi^+ \psi') = (3.2^{+1.8~+5.3}_{-0.9~-1.6}) \times 10^{-5}$, has also 
been obtained. 

\begin{figure}
\centering
\includegraphics[width=14cm]{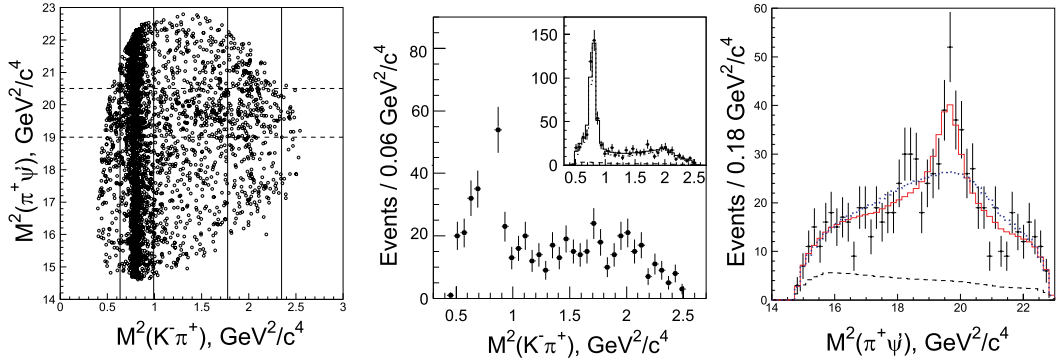}
\label{fig4}
\centering
\caption{ Left: The $B^0 \to K^{\mp} \pi^{\pm} \psi'$ Dalitz plot. 
Middle: A Dalitz plot projection for the horizontal slice
corresponding to the $Z(4430)$ region as a $K \pi$ invariant
mass distribution. 
Right: A Dalitz plot projection as a $\psi' \pi$ invariant
mass distribution with the $K^*$ veto applied. 
The dots with error bars represent data, the solid (dotted) 
histogram is the Dalitz plot fit result for the fit model with 
all $K\pi$ resonances and a single (without any) $\pi \psi'$ state, 
and  the dashed histogram represents the background.}
\end{figure}

\section{Summary}
 We have updated the analysis of $X(3872)$. There is no visible difference between the branching fractions of $B \to X(3872) K$ from charged and neutral $B$ mesons. We find no mass splitting in the $X(3872)$'s from these decays. The mass measured 
in the $D^0 \bar{D}^{*0}$ mode is compatible to that in the $J/\psi \pi^+ \pi^-$ mode. The mass of 
$X(3872)$ is still very close to the sum of $D^0$ and $D^{*0}$ masses. 
We confirm the signal of $Z(4430)$ with the similar significance as the previous analysis,
in Dalitz analysis taking into account several $K^*$ resonances. \\
\ \\

\end{document}